\newcommand{\beq}{\begin{equation}}
\newcommand{\eeq}{\end{equation}}
\newcommand{\bea}{\begin{eqnarray}}
\newcommand{\eea}{\end{eqnarray}}
\newcommand{\bear}{\begin{eqnarray*}}
\newcommand{\eear}{\end{eqnarray*}}

\newcommand{\rf}[1]{(\ref{#1})}

\documentclass[pre,showpacs,twocolumn,superscriptaddress,showkeys]{revtex4}

\usepackage{epsfig}
\usepackage{psfrag}
\usepackage{graphicx}

\begin{document}
\title
{Directed abelian algebras and their applications to stochastic models}
\author{F. C. \surname{Alcaraz}}
\email{alcaraz@if.sc.usp.br}
\affiliation{Instituto de F\'{\i}sica de S\~ao Carlos, Universidade de S\~ao 
Paulo, \\
Caixa Postal 369, 13560-590, S\~ao Carlos, S\~ao Paulo, Brazil. 
\vspace{0.1cm}}
\author{V. \surname{Rittenberg}}
\email{vladimir@th.physik.uni-bonn.de}
\affiliation{Physikalisches Institut, Bonn University, 53115 Bonn, Germany} 

\date{\today}

\begin{abstract}
To each directed acyclic graph (this includes some D-dimensional lattices) 
one can associate some abelian algebras that we call directed abelian 
algebras (DAA). On each site of the graph one attaches a generator of 
the algebra. These algebras depend on several parameters and  are 
semisimple.  Using any DAA one 
can define a family of Hamiltonians which give the continuous time 
evolution of a stochastic process. The calculation of the spectra and 
ground state wavefunctions (stationary states probability distributions) 
is an easy algebraic exercise.
 If one considers D-dimensional lattices and choose Hamiltonians linear in 
the generators, in the finite-size scaling the Hamiltonian spectrum is  
gapless with a critical dynamic exponent z = D.
 One possible application of the DAA is to sandpile models. 
In the paper we present 
this application  considering one and two dimensional 
lattices. In the one dimensional case, when the DAA conserves the number 
of particles, the avalanches belong to the random walker universality 
class (critical exponent $\sigma_{\tau} = 3/2$). We study the local density of particles inside large avalanches 
showing a depletion of particles at the source of the avalanche and 
an enrichment at its end. In two dimensions we did extensive Monte-Carlo 
simulations and found $\sigma_{\tau} = 1.782 \pm 0.005$.
\end{abstract}

\maketitle
\section{ Introduction} \label{sect1}

It was Dhar \cite{D1,D2} who pointed out that certain abelian algebras play an 
important role in understanding the properties of sandpile models. 
The time 
evolution of these systems is given, in the continuum limit, 
by a nonhermitian 
Hamiltonian  which acts nonlocally in the configuration space.

 The expression of the Hamiltonian contains local generators of an abelian 
algebra acting in the vector space of the regular representation of the 
abelian algebra. In some cases this vector space can be  
identified with the 
recurrence states (configurations) of the sandpile models.
 These algebras have a special structure describing topplings from a 
site to its neighbors. The net result being a highly nonlocal 
interaction. 

The abelian algebra appears in two different ways. On the one hand 
it determines the Hamiltonian which drives the system to the stationary 
state. On the other hand, if one considers the 
action of one of the generators on the stationary state of the system, the 
algebra describes through topplings a "short" (instantaneous) time behavior 
of an avalanche triggered by the local action of the generator. One can attach
a virtual time (this is the "short" time) to the evolution of the avalanche.
This is possible due to the abelian nature of the algebras. In
some cases the real and virtual times can get combined in a unique time (see 
\cite{TP} for an example), we are not considering  this case here.

 In the present paper we are going to present a large class of 
abelian algebras that we  call directed abelian algebras (DAA). 
They are defined as follows. One takes a directed acyclic graph \cite{FH}
 (these 
graphs are much used in mathematics and computer sciences \cite{DAG}). 
They are 
formed by oriented links (edges) to which we attach arrows. There are no oriented 
closed loops.  To each 
site (vertex) of that graph  one attaches a 
generator of DAA. There are as many relations among the generators as 
sites in the  
 graph. Examples of directed acyclic 
graphs are some lattices in any number of dimensions.
 The direction  in which  the sandpiles propagate
 can be identified with the virtual time. 
The evolution of the avalanches can be seen as a growth process in one 
space-dimension less \cite{CN}.

 The DAA are semisimple (all representations are decomposable into irreducible 
representations). They also have a special property: they have only one-dimensional 
irreducible representations and the number of nonequivalent representations coincides with 
the dimension of the regular 
representation. Since the Hamiltonians act onto the regular representations, they can be 
diagonalized in a trivial way. This stays true even if the Hamiltonians describe the 
evolution of  systems in any number of dimensions. 
At least in some cases, all the 
eigenfunctions can be easily obtained.

The stationary state, which is of product measure form (no correlations!),
 can also
be obtained in any number $D$ of dimensions. The thermodynamic limit of the spectra is also very simple.  

 All systems described by directed algebras are critical with the dynamic  
critical exponent having the value $z = D$.  This does not imply that the avalanches have to be critical since they evolve in virtual time. They 
might have exponential tails.

The paper is organized as follows. In Section 2 we define a quadratic DAA on a 
one-dimensional lattice.
   It is a two-state model in which one has a particle or a vacancy on 
each site.  We give the Hamiltonian which gives the time evolution of this  system and 
show how to compute exactly its spectrum.  The 
ground-state wavefunction  is of product form. The average density is uniform in spite of having an open system.

 The Hamiltonian spectrum and eigenfunctions are given in Appendix A for the 
physically relevant case in which the number of particles is conserved. 

 In Section 3 we study the avalanches in the model defined in Section 2. We show that 
they belong to the "random walker" universality class \cite{MZ}. 
The probability to find an 
avalanche of virtual time $T$ corresponds to the probability to have a 
first passage at time $T$
 of the random walker.
The connection between the present paper and the totally asymmetric
Oslo model \cite{GP,SC} is presented in Appendix B.
We also look at the average density of particles left in the system when the avalanche 
stopped at time $T$. The density of particles is not uniform anymore.
 The physical process 
being directed, the region near the beginning of the avalanche is depleted and the region
at the end of the avalanche is enriched in particles. For long avalanches, the particle density distribution functions  
become singular at both ends of the avalanches. It is easy to understand the physical 
origin of the two singularities.

 Higher order one-dimensional DAA are defined in Section 4. They correspond to N-state 
models. All the properties of the 2-state models are recovered and no new qualitative 
features appear. 

The construction of a DAA on an arbitrary directed acyclic graph is 
presented in Section 5. As applications we consider   DAA on a  
one-dimensional lattice, in which a site $i$ is related to two successive 
sites $i + 1$ and $i + 2$, and on two-dimensional lattices. 
 In two dimensions, one obtains avalanches belonging to the universality class described 
in \cite{RV,PB,KMY}. 
We have done large scale Monte-Carlo simulations to measure the
exponent $\sigma_{\tau}$ (usually denoted by $\tau_t$). 
The result doesn't agree with the expected value $\sigma_{\tau}= 7/4$.
 
Our conclusions are presented in Section 6.

\section{ A quadratic directed abelian algebra and its application to a 
stochastic process.}\label{sect2}

In this section we first show how to define the Hamiltonian describing a 
stochastic process by using an associative algebra. We then consider a 
simple application of quadratic directed abelian algebra defined on a one-dimensional 
lattice.  Associated to this algebra we may define a family of stochastic 
processes, sharing the same stationary state.  

 Let us now consider an associative algebra with generators $A(i)$
($i=1,2,\ldots,m$). Taking products of the generators one gets the words $W(
r)$.
The algebra is defined by giving some relations between the words. If we
can choose the linearly independent words $\{W(r); r=1,\ldots,n_W\}$ such that any product of them  verify the
relation:
\begin{equation} \label{1}
W(r)W(s) =  \sum_{q=1}^{n_W} p_q^{r,s}W(q),\;\;\;\;\;  p_q^{r,s}\geq 0,\;\;\;\;\sum_{q=1}^{n_W}
 p_q^{r,s} = 1,
\end{equation}
then the Hamiltonian
\begin{equation} \label{2}
H = \sum_{u=1}^{n_W} c(u) (1 - W(u)),
\end{equation}
acting in the vector space
defined by the basis of independent  words $\{W(s)\}$,  is an intensity matrix
provided that  the coefficients $c(u)$ are nonnegative \cite{AR}. $H$ is acting from 
the left on 
the
words $W(s)$ of the vector space: $W(u)W(s)$ ($W(u)$ is one on the terms in
\rf{2} and $W(s)$ belongs to the vector space). The action of the $n_W$  
independent
words $\{W(u)\}$ on the vector space defined by the same words gives the regular
representation of the algebra.

If the algebra $A$ contains a
left ideal \cite{footnote}
 defined by the words
 $I$, the Hamiltonian \rf{2} acting
on this ideal gives again a stochastic process. If $A$ has several ideals
$I_1 \subset I_2 \subset \cdots \subset I_n$, $H$ has a block triangular form. The ground-state
wavefunction (eigenvalue equal to zero) is a linear combination of the states (words) which
define the vector space $I_1$. These states correspond to the  recurrent
states. The words belonging to the ideals $I_2,\ldots,I_n$ but not belonging
 to
$I_1$ do not appear in the stationary state.

One can obtain Hamiltonians describing stochastic processes using
associative algebras even if the positivity conditions \rf{1} are not
fulfilled. This can be achieved by not working in the regular
representation of the algebra but in other representations. The following
simple example illustrates the procedure.

 Consider the quadratic algebra with only one generator $a$:
\beq \label{a1}
a^2 = (1-\mu)a + \mu e,
\quad e^2 = e, \quad ea = a e     
\eeq
and
\beq \label{a2}
H = 1-a,
\eeq
$\mu$ is a parameter. If $0 \leq \mu \leq 1$, the condition \rf{1}
 is satisfied. H is an
intensity matrix (order $e$ and $a$):
\bea \label{a3} 
H = &&\left(\matrix{ 0 & -\mu \cr 0 & \mu}\right),
\eea
where $e$ is the $2\times 2$ unit matrix.
Let us now perform the similarity transformation:
\beq \label{a4}
H' = S H S^{-1}                                          (C)
\eeq
where
\bea
&&S = \left(\matrix{1 & \xi \cr \xi & 1}\right) \nonumber
\eea
and $\xi = (1 - \alpha) /(1 - \beta)$. $H'$ is:
\bea \label{a5}
&&H' =  \left(\matrix{\alpha & -\beta\cr -\alpha & \beta}\right).
\eea                                      
The algebra \rf{a1} stays unchanged with $\mu = \alpha + \beta - 1$. 
The matrix \rf{a5}
is the most general $2\times 2$ intensity matrix ( $0\leq \alpha$, 
$\beta \leq 1$ ). Notice
that the physical process depends on two parameters instead of one (which
appears in the algebra \rf{a1}) and that the range of $\mu$ has changed: 
$- 1 < \mu \leq 1$.

 The probabilities $P_q(t)$ to find the words $W(q)$ at the time t can be 
obtained from the master equation
\begin{equation}\label{3}
\frac{d}{dt} P(t) = -H P(t),
\end{equation}
where
\begin{equation}\label{4}
P(t) = \sum_{q=1}^{n_W} P_q(t)W(q).
\end{equation}


 We now apply this formalism to  a simple example of a quadratic DAA 
(for applications to non-abelian algebras, see \cite{AR}). Consider 
a one-dimensional lattice with $L$ sites. We attach to each site $i$ ($i = 
1,2,\ldots,L$) a generator $a_i$. A quadratic DAA is defined by the 
relations:
\beq \label{5}
a_i^2 = p_1a_{i+1}^2+p_2a_{i+1}+p_3+q_1a_ia_{i+1}+q_2a_i, 
\eeq
for  $i=1,2,\ldots,L-1$, 
\beq \label{6}
a_L^2 = \mu + (1-\mu)a_L,
\eeq
\beq \label{7}
[a_i,a_j]=0, \quad i,j=1,2,\ldots,L,
\eeq
where the $p_m$ and $q_n$ are probabilities and
\beq \label{8}
\mu = p_1+p_2+p_3, \quad 1-\mu = q_1+q_2.
\eeq
 Notice that the words with a given value $i$ are related to words with  
values  $j>i$, therefore we have a preferred direction on the lattice.
The relation \rf{6} was obtained taking $a_{L+1} = 1$ in equation \rf{5}.

 The $2^L$ monomials $1, a_1,\ldots,a_L, 
a_1a_2,\ldots,a_{L-1}a_L,\ldots,$ $a_1a_2\cdots a_L$ are the independent words of the DAA (they 
define the regular representations of the DAA). One can check, 
using \rf{5}-\rf{8}, that the relations \rf{1} are verified. 
The DAA \rf{5}-\rf{7} can 
therefore be used to define a family of stochastic processes  whose 
Hamiltonians are obtained by arbitrary choices of $c(u)$ ($u=1,\ldots,2^L$)
in \rf{2}.
 If $\mu = 0$, the algebra \rf{5}-\rf{8} has an ideal
\beq \label{9}
  I_1 = a_1a_2\cdots a_L,
\eeq
which gives in the regular representation an absorbing state in the stochastic process defined by
\rf{2}. 

 One can map the vector space of the regular representation onto a lattice 
with $L$ sites. On each site $i$ one has a particle if $a_i$ appears in 
the 
monomial or a vacancy if $a_i$ doesn't appear in the monomial. 
To the word "1" 
one associates the empty lattice. There are $2^L$ configurations 
obtained in 
this way.

 The $a_i$ ($i = 1,2,\dots,L$) acting onto the regular representation 
are not in the 
diagonal form. They can however be diagonalized simultaneously and one 
obtains in this way   
 $2^L$ one-dimensional irreducible representations of the algebra. The 
eigenvalues of $a_i$ denoted by $x_i$ can be obtained 
recursively using the distinct eigenvalues $x_{i + 1}$ of $a_{i + 1}$ 
\bea \label{10}
&&x_L = 1,\mu, \nonumber \\
&&x_i^2-x_i(q_1x_{i+1}+q_2)-(p_1x_{i+1}^2+p_2x_{i+1}+p_3) = 0,\nonumber  
\eea
where $i=L-1,L-2,\ldots,1$.

Notice that one has to solve only quadratic equations and that the $a_i$ 
have a common eigenvalue $x_i = 1$. One can easily find the corresponding
eigenfunction in the regular representation:
\beq \label{11}
a_i \Phi = \Phi, \quad i=1,2,\ldots,L.
\eeq
 The expression of $\Phi$ is:
\beq \label{12}
\Phi = \prod_{i=1}^L \frac{\mu +a_i}{1+\mu}.
\eeq
  In the expression of $\Phi$ enters only the parameter $\mu$
 instead 
of the four parameters appearing in \rf{5}. The proof is straightforward. 
We 
have:
\beq \label{13}
a_L\frac{\mu +a_L}{1+\mu}= \frac{\mu +a_L}{1+\mu}
\eeq
Where we have used \rf{6}. We now use \rf{5} and \rf{13} to obtain:
\beq \label{14} 
a_{L-1}\frac{\mu+a_{L-1}}{1+\mu}\frac{\mu +a_L}{1+\mu} = \frac{\mu+a_{L-1}}{1+\mu}\frac{\mu +a_L}{1+\mu} 
\eeq
and repeat the procedure to get, due to \rf{7},
\beq \label{15}
a_i\Phi = \Phi, \quad i=L-2,\ldots,1.
\eeq
The eigenstate $\Phi$, due to \rf{10}, gives the stationary state for 
the arbitrary stochastic models defined by \rf{2}, i. e. $H\Phi =0$.
 The average density in the stationary state \rf{12} is equal to 
$1/(1 + \mu)$
and varies between $1/2$ and $1$.
 
Other equivalent representations of the algebra \rf{5}
 can be used
to describe  stochastic processes. In Appendix B we give two examples. One
of them is the totally asymmetric Oslo model \cite{GP,SC}.

 We now consider the simple  stochastic process given by the Hamiltonian:
\beq \label{16}
H = \sum_{i=1}^L(1-a_i)/L.
\eeq
 The physical interpretation of the stochastic process described by \rf{16} 
can be more easily  understood using the configuration space and discrete times. 
At the time $t$ one takes a configuration with a certain distribution of 
particles on the $L$ sites (not more than one per site). With a probability 
$1/L$ a particle is added at the site $i$. If the site is empty, the particle 
stays at $i$. If one has a particle at $i$ already, one obtains several 
configurations with a probability given by the toppling rules given by the 
DAA \rf{5}.

 Notice that the stationary state  is of product form and therefore there are no 
correlations in the stationary state.

 To compute the spectrum of $H$, we can take all the $a_i$ in their 
diagonal representation. Since $H$ acts in the regular representation, all 
the irreducible one-dimensional representations of the DAA have to appear. 
  The number of irreps coincides  with the dimension of the vector space given 
by the regular representation. Therefore, each representation appears only once and the 
calculation of the spectrum is trivial (see Appendix A). It is interesting to compute the
energy gap of $H$. The ground-state energy is zero. The first excited state 
is obtained taking $a_i  = 1$ ($i = L, L-1,...,2$). It follows from \rf{5}
 that 
$a_1$ can take two values: $1$ and $-\mu$. One obtains two one-dimensional 
representations. The first one is the ground state with energy zero and  the second one, corresponding to the first excited state with 
energy $E_1 = (1 + \mu)/L$ (we call energies the eigenvalues of $H$). 
The dynamic critical exponent $z$ follows:
\beq \label{18}
\lim_{L\to \infty} LE_1 = 1+\mu,
\eeq  
therefore $z = 1$. 
This result is interesting for several reasons. The value 
$z = 1$ is also obtained when the system is conformal invariant \cite{AR}. 
This 
is certainly not the case here since there are no correlations in the 
stationary state. The finite-size scaling limit of the spectra can be 
changed (through the change of the parameters in \rf{5}) almost at will and 
are not described by representations of the Virasoro algebra as expected 
in a conformal invariant theory. The result is not surprising since in the 
configuration space, each $a_i$ acts through topplings in a highly nonlocal 
way and there is no reason to expect conformal invariance. It
  is nice 
to have an example which illustrates this phenomenon in full generality.

 There is another interesting consequence of \rf{18}. Since the energy gap   in the thermodynamic limit  
vanishes, the system is critical. Does this imply that the avalanches, to 
be discussed later, are described always by power laws as expected by SOC? The
answer is no. As one can see from \rf{5}, unless all the coefficients 
$p_2, p_3$ 
and $q_2$ vanish, the number of particles is not conserved, 
particles are lost. One   expects therefore an
exponential decay for the probability of having  an avalanche of a certain 
size \cite{SR,TT}.
  For avalanches which are critical one defines a dynamic critical 
exponent $z_a$ \cite{BVZ} related to the virtual time in which the avalanche 
occurs. It is not clear to us why the two exponents $z$ and $z_a$
 should coincide
although in many examples they do.

 The spectrum and eigenfunctions of $H$ are given in Appendix A for the 
simpler case in which one conserves the number of particles in \rf{5}. The 
knowledge of the spectra allows us to get information of time dependent 
quantities. If one starts, for example, 
 with an empty lattice, one can look 
at the time dependence of the average density which, in the large $L$ limit,
should reach algebraically its value in the stationary state.

 There is a purely algebraic way to compute the density for finite $L$. 
Let us
assume that at $t = 0$ we start with an empty lattice. This corresponds to the
word "1" in the algebra. In the basis of monomials (see \rf{4}), one has
\beq \label{19}
P(t) = e^{-Ht}\cdot 1 = e^{-Ht}.
\eeq

 After expanding the exponential in \rf{19} we use  the algebra \rf{5}
  and expresses  $P(t)$ in terms of monomials. We take 
$a_i = y$
(for all $i$'s) and obtain the function $P(t,y)$. 
The density is given by the 
expression:
\beq \label{20}
\rho(t) = \frac{dP(t,y)}{dy}{\Bigg |}_{y=1}.
\eeq

 Let us observe that  we can take the parameters in \rf{5} and \rf{6} $i$-dependent, 
the ground-state wavefunction maintaining  the product form:
\beq \label{21}
\Phi = \prod_{i=1}^L \frac{\mu_i +a_i}{1+\mu_i},
\eeq
where  the definition of $\mu_i$ is obvious. 
The dynamic critical exponent 
$z$ stays unchanged.

\section{ A quadratic directed abelian algebra and avalanches.}\label{sect3a}

 In the last section we have discussed the time evolution of a system described by $H$ given
by \rf{16}. In order to study its spectrum and obtain the ground-state wavefunction we have used 
the quadratic DAA \rf{5}-\rf{8}. From these results, in this section, we need only the 
expression \rf{20} and use the DAA to study avalanches. To simplify the calculation we 
consider the case in which the number of particles is conserved. In this 
case, relations \rf{5} take the form:

\beq \label{22}
a_i^2 = \mu a_{i+1}^2 +(1-\mu)a_ia_{i+1}.
\eeq

 The physical meaning of \rf{22} is simple: if one has two particles on the site $i$, with a 
probability $\mu$ they reach the site $i + 1$ and with a probability 
($1 - \mu$) one particle moves 
to the site $i +  1$  and one stays at the site $i$. Multiplying \rf{22} by a power of $a_i$ we can 
find what are the probabilities that n particles on site $i$ move in block on the site $i + 1$ 
or only $n-1$ particles move and one stays at the site $i$. 
Obviously the number of particles is not conserved at 
the end of the system.

 To examine avalanches in our system, we take 
 the stationary state given by \rf{12} and consider only the 
configurations in which the first site is occupied (they have a 
probability $1/(1 + \mu)$). We add one particle on the first site  and 
analyze the
consequences. The overall effect of adding the particle is trivial to 
derive:
\beq \label{23} 
a_1^2 \prod_{i=2}^L \frac{\mu+a_i}{1+\mu} = 
(\mu +(1-\mu)a_1)\prod_{i=2}^L\frac{\mu+a_i}{1+\mu},
\eeq
where we have used \rf{11} and \rf{12}.
This tells us that the changes take place only at the first site. With 
probability $\mu$ it becomes empty and with probability $(1-\mu)$ is 
occupied by a particle.
 
We are interested in  more detailed 
properties of this process. First we would like to know what is the 
probability $p(k)$ to have the first $k$ sites affected by the action of 
$a_1^2$
and the remaining $L-k$ sites untouched (this defines an avalanche of size 
$k$, starting at the first site). The physics inside an avalanche is easy to understand looking at 
\rf{22}.
 Particles close to the origin of the avalanche are pushed towards its end. We
will be interested therefore to obtain  the non-homogeneous density of
particles inside an avalanche. Let us first obtain $p(k)$. 

 We use \rf{22} and obtain:
\bea \label{24}
&&a_1^2 \prod_{i=2}^L \frac{\mu+a_i}{1+\mu} =
\left[\frac{\mu(1-\mu)a_1a_2}{1+\mu} \right .\nonumber \\
&&\left. + 
\frac{\mu a_2^3+\mu^2a_2^2+(1-\mu)a_1a_2^2}{1+\mu}\right]
\prod_{i=3}^L \frac{\mu +a_i}{1+\mu}.
\eea
 One can see that we have gotten an avalanche given by $a_1a_2$ of size 2 
(the first two sites occupied) with probability $\mu(1-\mu)/(1+\mu)$. 
If we are not interested in the content of the avalanche, and only on its 
size, we can formally take $a_1=1$ in \rf{24}  to get:
\bea \label{24a}
&&a_1^2 \prod_{i=2}^L \frac{\mu+a_i}{1+\mu} {\hat{=}} 
\frac{\mu(1-\mu)a_2+(1-\mu+\mu^2)a_2^2+\mu a_2^3}{1+\mu} \nonumber \\
&&\times \prod_{i=3}^L\frac{\mu+a_i}{1+\mu}.
\eea

 If we want 
to proceed further, and compute the probabilities of larger avalanches we have to compute quantities like:
\bea \label{25}
a_i^n \frac{\mu+a_i}{1+\mu} &=& \frac{1}{1+\mu}[B_0^{(n)}a_{i+1}^{n+1} + 
(C_1^{(n)}a_i +C_0^{(n)})a_{i+1}^n \nonumber \\
&&+D_1^{(n)}a_ia_{i+1}^{n-1}],
\eea
in which we have taken into account the fact that the algebra \rf{22} 
conserves the number of particles. Multiplying \rf{25} by $a_i$ one obtains 
the recurrence relations:
\bea \label{26}
&&B_0^{(n+1)} = \mu C_1^{(n)}; \; C_1^{(n+1)}=B_0^{(n)}+(1-\mu)C_1^{(n)} 
\nonumber \\
&&C_0^{(n+1)} = \mu D_1^{(n)}; \; D_1^{(n+1)}=C_0^{(n)}+(1-\mu)D_1^{(n)} 
\eea
with
\bea \label{27}
&&C_1^{(0)} =1, \quad D_1^{(0)} = 0, \; C_1^{(1)} = 1-\mu, \;
 D_1^{(1)}=\mu
\eea   
From which we get:
\beq \label{28}
D_1^{(n)} = \mu C_1^{(n-1)}
\eeq
and 
\beq \label{29}
C_1^{(n+1)} -C_1^{(n)} = -\mu(C_1^{(n)}-C_1^{(n-1)}).
\eeq

 The solution of the recurrence relation \rf{29} is
\beq \label{30}
C_1^{(n)} = \frac{1- (-\mu)^{n+1}}{1+\mu}.
\eeq

In order to look for avalanches, in \rf{25} one takes $a_i = 1$ and using 
\rf{26} 
and \rf{28}, one obtains
\bea \label{31}
&&a_i^n \frac{\mu+a_i}{1+\mu} {\hat{=}} 
\frac{1}{1+\mu}[\mu C_1^{(n-1)}a_{i+1}^{n+1} \nonumber \\+
&&(C_1^{(n)}+\mu^2C_1^{(n-2)})a_{i+1}^n 
+ 
\mu C_1^{(n-1)}a_{i+1}^{n-1}].
\eea

 If $\mu \neq 1$  (for $\mu = 1$ the process is deterministic and the two 
particles on site 1 move on site $L$ where they leave the system) from 
\rf{30} 
one can see that for large values of $n$, $C_1^{(n)}$ 
becomes independent on $n$ 
and equal to $1/(1+\mu)$. Substituting this result in \rf{31} one obtains:
\beq \label{32}
a_i^n \frac{\mu+a_i}{1+\mu} \hat{=} \frac{1}{(1+\mu)^2} 
[\mu a_{i+1}^{n+1} + (1+\mu^2)a_{i+1}^n+\mu a_{i+1}^{n-1}].
\eeq

 One could have obtained this last equation without solving the recurrence relation 
\rf{27}-\rf{29}
(we are going to use this method in the next section).
Equation \rf{29} has an $n$-independent solution $C_1^{(n)} = C$. 
We introduce this
solution in \rf{31} and ask for the sum of the coefficients to be equal to 
 1. This
gives $C = 1/(1 + \mu)$.

The whole virtual time evolution ($\tau \geq 1$), of the avalanches 
is given by   the following  expression
\beq \label{q3}
a_1^2\prod_{i=2}^L\frac{\mu+a_i}{1+\mu} \hat{=} 
\sum_{n=1}^{\tau}P_n(\tau)a_{\tau}^n\prod_{j=\tau+1}^L\frac{\mu+a_j}{1+\mu},\eeq
where $P_n(\tau)$ is the probability that at virtual time $\tau$ an 
avalanche,  with $n$ particles  at site $i=\tau $, is taking place. 
These probabilities, 
from \rf{22} and \rf{31}, satisfy the recurrence relations
\bea \label{d1}
P_1(\tau) &=& P_2(\tau-1)R_-^{(2)}, \nonumber \\
P_2(\tau) &=& P_2(\tau-1)R_0^{(2)}+P_3(\tau -1)R_-^{(3)}, \nonumber \\
P_n(\tau) &=& P_{n+1}(\tau-1)R_-^{(n+1)} + P_n(\tau-1)R_0^{(n)} 
\nonumber \\ 
&+& P_{n-1}(\tau-1)R_+^{(n-1)}, 
\eea
for $2\leq n\leq \tau$, with 
\beq \label{d2}
P_n(1) = \delta_{n,2},
\eeq
and
\bea \label{q2}
&&R_+^{(n)}=\frac{f^{(n-1)}}{1+\mu}, R_-^{(n)} = \mu \frac{1-f^{(n-1)}}{1+\mu}, 
\nonumber \\
&& R_0^{(n)}=1-R_+^{(n)}-R_-^{(n)}, f^{(n)} = 
\frac{\mu + (-\mu)^{n+1}}{1+\mu}.
\eea 

 We can visualize these last  equations  in the following way. 
We take two coordinates: the 
virtual time $\tau$ (discritized values) as the horizontal axis and a  coordinate $n$
(discritized values) as a vertical axis. Equation \rf{q3} 
describes a random walker which 
at time $\tau$ 
is at the position $n$, and 
moves to the positions $n+1$, $n-1$ or stays at $n$ with probabilitoes 
which are $n$-dependent.
 The probability of having  an avalanche of size $k$ is given by  the first 
passage at virtual time $\tau = T=k$ to return to its initial 
position $x=1$, and is given by $p(\tau)=P_1(\tau)$. This probability 
is not simple to calculate due to the $n$-dependence of the 
random walk parameters $R_-^{(n)}$, $R_0^{(n)}$ and $R_+^{(n)}$ in \rf{d1}. 
However the large size avalanches are dominated by  random walks that 
stay mostly   at large distances from the origin ($n$ large). 
We take $f^{(n)}=\mu/(1+\mu)$ in \rf{q2} and obtain
\beq \label{d3}
R_+^{(n)} = R_-^{(n)}=R=\mu/(1+\mu)^2, R_0^{(n)}= 1 -2R=
\frac{1+\mu^2}{(1+\mu)^2}.
\eeq
 With  probabilities $\mu/(1 + \mu)^2$, the walker  moves to $n + 1$ or $n - 1$ 
and with a probability $(1 + \mu^2)/(1 + \mu)^2$ stays at $n$.   
In the continuum, one obtains the diffusion equation with a diffusion
constant $D = \mu/(1 +\mu)^2$. Consequently, the probability of having 
large avalanches of size $k$ corresponds to the usual first passage time 
at $\tau=T=k$ \cite{SR}
\beq \label{33a}
 p(T) =P_1(T) \sim  \frac{1}{\sqrt{DT^3}}.
\eeq
If one defines the exponent $\sigma_{\tau}$ by the relation $p(T) \approx1/T^{\sigma_{\tau}}$, where $T$ is the 
duration of the avalanche, one obtains  $\sigma_{\tau} = 3/2$.
The approximation \rf{d3} for the probabilities \rf{q2} is valid if
$n>>1/|\ln \mu|$. Since $n$ is of the order of $\sqrt{T}$, 
we conclude that one can
see the avalanches described by the power law \rf{33a} for
\beq \label{e1}
T >>T_c =1/(\ln \mu)^2.
\eeq

\begin{figure}[ht!]
\centering
{\includegraphics[angle=0,scale=0.46]{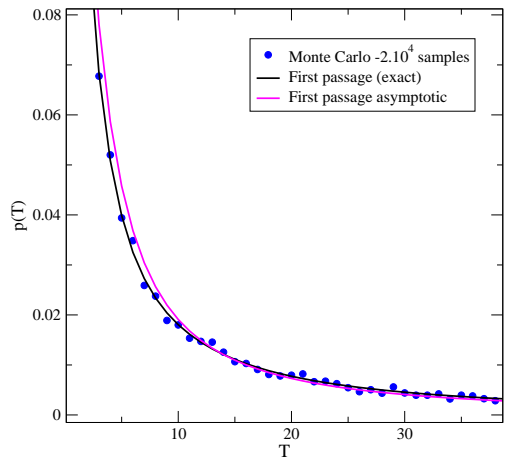}}
\caption{(Color online) Exact and Monte Carlo results for the probability distribution of avalanches of size $T$ for the stochastic model defined by 
\rf{16} and \rf{22}, with $\mu = 1/4$. The dots are the results from Monte Carlo simulations. 
The black curve is the exact result, calculated algebraically. 
The red curve is obtained by considering the asymptotic behavior of the
 recurrence relations \rf{d3} in \rf{d1}.}
\label{fig1}
\end{figure}

In order to illustrate the approximation \rf{d3} we compare in Fig.~\ref{fig1} the results obtained by the exact recurrence relations with 
$\mu=1/4$  \rf{d1} (black 
curve) with the one obtained by using the asymptotic relations \rf{d3} (red curve). We only show the probabilities for relatively small avalanches 
($\tau < 50$) where the curves show a more pronounced difference 
($T_c = 1$ in this case). We also 
show (blue dots), for comparison, the average results obtained from  
$2\times 10^4$ samples 
in a Monte Carlo simulation.

 Having in mind applications to two-dimensional systems (see Section 5) we
are interested to know how large have to be the avalanches to allow us to
get a good estimate for $\sigma_{\tau}$. We consider the quantity:
\beq \label{est1}
\sigma_{\tau}(T) = T\frac{p(T) - p(T+1)}{p(T)},
\eeq
which is an estimate for $\sigma_{\tau}$. In Fig.~\rf{pp2}
 we show the values of $\sigma_{\tau}(T)$ 
for $\mu = 0.99$ ($T_c = 9900$ in this case). We observe  that $\sigma_{\tau}(T)$
 first
increases with $T$ giving a maximum value of 1.91 and then decreases. Notice
that even taking very large avalanches one still gets a poor estimate:
$\sigma_{\tau}(10^6) = 1.5445$. This estimate is larger than the expected value 1.5. Using extrapolation techniques (8-degree polynomial fitting) we can improve substantially the estimate and get $\sigma_{\tau} = 1.5006\pm0.0008$. 
We have to stress that for $p(T)$ we have used exact results and not Monte Carlo
simulations as in Section 5.
%
%
\begin{figure}[ht!]
\centering
{\includegraphics[angle=0,scale=0.46]{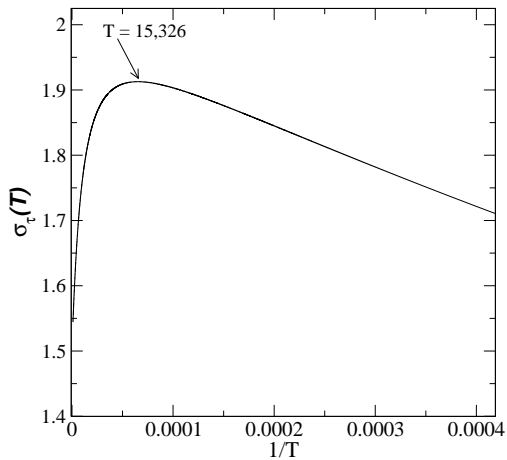}}
\caption{The estimate $\sigma_{\tau}(T)$ defined in \rf{est1} for
$\mu = 0.99$ and $2400<T<10^6$. The estimate increases, with $T$, up tp
$T=15,456$. The last value of the estimate is $\sigma_{\tau}(10^6)=1.5445$.}
\label{pp2}
\end{figure}
%
%
\begin{figure}[ht!]
\centering
{\includegraphics[angle=0,scale=0.46]{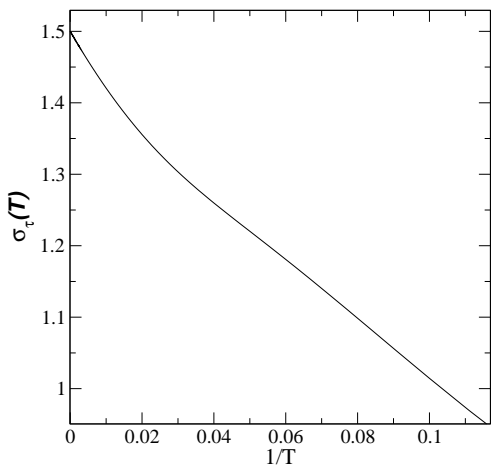}}
\caption{The estimate $\sigma_{\tau}(T)$ defined in \rf{est1} for
$\mu = 0.1$ and $1<T<10^6$. The last value of the estimate is $\sigma_{\tau}(10^6) =
1.49999$.}
\label{pp3}
\end{figure}

 If $\mu$ is small, the estimates of $\sigma_{\tau}(T)$ improve dramatically. In 
Fig.~\rf{pp3} 
we show $\sigma_{\tau}(T)$ for $\mu = 0.1$ ($T_c$ is negligible in this case). 

We
observe that in contrast with the case $\mu = 0.99$, the estimates increase
with $T$ and one can get a very good precision for relatively small values
of $T$. A best fit to the data gives :
\beq \label{estt}
 p(T) = A/T^g + B/T^h,
\eeq     
where $A = 1.0794 \pm 0.0003$,$ B = - 7.16\pm 0.05$, $g = 1.50000 \pm 0.00001$, $h =
2.448 \pm 0.003$.

In this way one has  obtained  a concrete realization of the model 
of 
 Maslov and Zhang  \cite{MZ} in which $z_a = 1$.
The same result was obtained   in a different context. 
A generalized totally asymmetric hopping processes on a ring (fixed density of particles) was considered \cite{PPH}. More than one particle per site 
are allowed. 
 The rules of the hopping processes 
are fixed by the condition of integrability of the system (one can write the Bethe 
ansatz). Interestingly enough the rules coincide with those obtained from the DAA \rf{22}. 
There is no relation 
\rf{6} on a ring. If the density of particles on the ring is equal to 
the average density $1/(1+\mu)$ obtained in our model, the two models describe the same 
large avalanches.

 We would like to make the connection between our approach and the work of
Stapleton and Christensen on the totally asymmetric Oslo model (TAOM) and
its generalizations \cite{SC}, where similar results were obtained. 
As shown in
detail in Appendix B in the case of the two-state model, the TAOM
corresponds to take another representation (not the regular one) of the
same algebra \rf{22}. This representation depends on two parameters
whereas the algebra depends only on one. The same picture stays valid
for the multi-state one-dimensional models discussed in Section 4. We have
directly 
proven the correspondence to the random walker process and
proceeded by deriving the probability distribution of avalanches. In \cite{SC}
the authors have computed moments of the toppling probability distribution
and came to the same conclusion.

 Up to now we were interested in the probability of having an avalanche of 
size $T$. We would like to know what is the physics of a given avalanche. 
It is useful to visualize the virtual time $\tau$ either as a time 
coordinate or a space coordinate.

\begin{figure}[ht!]
\centering
{\includegraphics[angle=0,scale=0.46]{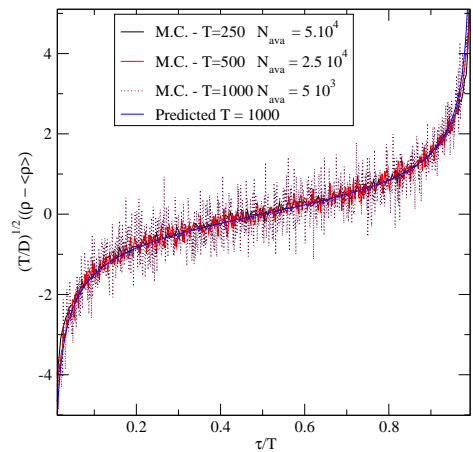}}
\caption{(Color online) Depletion and enrichment of particles in avalanches of different 
sizes $T$. $(T/D)^{1/2} (\rho(\tau,T) -<\rho>)$ as a function of $\tau/T$
 obtained from 
Monte-Carlo simulations for $T = 250, 500$ and $1000$. The numbers of 
avalanches $N_{ava}$ observed in each case are given in the insert. The 
solid curve is the result of the calculation using \rf{d3p} and \rf{d4p} for $T = 1000$.}
\label{fig1a}
\end{figure}

Looking at the DAA \rf{22} we suspect that particles 
present in the 
stationary state (average density $<\rho> = 1/(1 + \mu)$) 
are pushed away from 
the origin of the avalanche towards its end. For a given avalanche of size 
$T$ we consider the quantity : $\rho(\tau, T) - <\rho>$, 
where $\rho(\tau,T)$ is the avergae density at virtual time $\tau$.
 In order to 
compute this quantity we could use \rf{25}
 without taking $a_i = 1$.  This is a lengthy calculation. Instead we did Monte-Carlo
simulations. In Fig.~\ref{fig1a}, for $\mu = 1/2$ we show ($\rho(\tau,T) - 
<\rho>$) 
multiplied by $\sqrt{T/D}$ as a function of $\tau/T$ 
for three avalanche sizes. 
The function is antisymmetric. $D$ is the diffusion constant 
($D = \mu/(1 + \mu)^2$). We notice not only a nice data collapse but a
singular depletion of particles for small values of $\tau/T$ and a singular
enrichment of particles for $\tau/T$ close to one. We have done similar 
simulations for other values of $\mu$ and the data come on top of those 
shown for $\mu = 1/2$. If one looks   $\tau$ as a 
space coordinate, one gets what one would expect to see in the aftermath 
of an avalanche.    

 For small values of $\tau/T$ one finds:
\bea \label{d1p}
&&\rho - <\rho> \sim -\frac{c}{4}\left(\frac{D}{\tau}\right)^{1/2}
\eea
where $c = 0.56$.

 Theoretically one can obtain the scaling function which describes the 
data shown in Fig.~\ref{fig1a} in the following way. 
We use the random walker 
approximation. Let $P(x,\tau;T) = P(x,T - \tau;T)$ be the conditional 
probability to find a random walker at time $0<\tau<T$ at the position $x$
 if 
the first passage time is $T$. This is a Brownian excursion \cite{MC}. 
The average position at time $\tau$ is
\beq \label{d2p}
<x>_{\tau,T} = \int_0^t xP(x,\tau;T)dx.
\eeq
 One can compute
\beq \label{d3p}
\rho(\tau,T) - <\rho> = -\frac{1}{2}\frac{d}{d\tau}<x>_{\tau,T}.
\eeq
 In order to do the calculation we have used the lattice expression 
$P(x,\tau;T)$ obtained in the simpler case in which the random walker
 which 
is at the position $x$ at time $\tau$ moves at $x + 1$ or $x - 1$ with 
probabilities $1/2$ at $\tau + 1$. 
The random walker starts at $x = 0$ at time 
$\tau = 0$ and returns at $x = 0$ at $\tau = T$ ($T$ even). 
One has \cite{AO}:
\bea \label{d4p}
&&P(x,\tau;T) = \frac{x^2T}{2}\frac{((T/2-1)!)^2}{(T-2)!} \nonumber \\
&& \times \frac{(\tau-1)!(T-\tau-1)!}{(\frac{\tau-x}{2})! 
(\frac{\tau+x}{2})!(\frac{T-\tau-x}{2})!(\frac{T-\tau+x}{2})!}. 
\eea

\begin{figure}[ht!]
\centering
{\includegraphics[angle=0,scale=0.46]{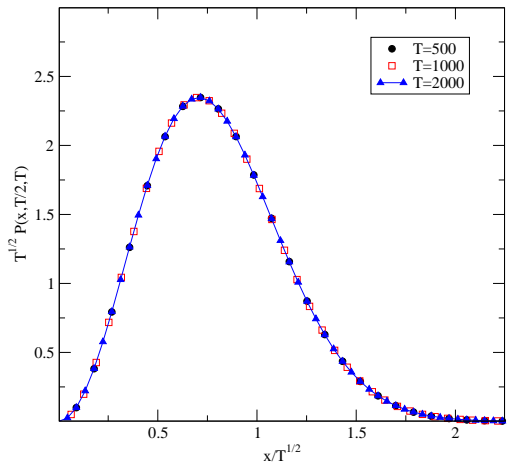}}
\caption{ (Color online) 
 The conditional probability function $P(x,T/2;T)$ multiplied by 
$T^{1/2}$ as a function of $x/T^{1/2}$ for $T = 500, 1000$ and $2000$.}
\label{fig1b}
\end{figure}

 Note that in the continuum the diffusion constant is $D = 1/2$ in this 
case. 
 The function $P(x,\tau;T)$ satisfies various scaling laws. 
In particular if 
we take $\tau = T/2$, it scales as shown in Fig.~\ref{fig1b}. A consequence of this 
scaling behavior is that 
\beq \label{d5p}
<x>_{\frac{T}{2},T} \sim T^{1/2}.
\eeq
For $\tau = T/2$, $<x>$ gets its maximum value. 
One has checked a more general 
scaling property:
\beq \label{d6p}
<x>_{\tau,T} = \sqrt{DT}f(\tau /T).
\eeq
Obviously $D = 1/2$ in the present case. 

 For small values of $\tau$, one gets:
\beq \label{d7p}
<x> = c\sqrt{D\tau},
\eeq
where $c =0.56\;$. The expression \rf{d7p} is what one expects for a 
random walker. 
The constant $c$ appears from the constraint $x > 0$. If we use \rf{d3p}
 and \rf{d7p} 
we  get \rf{d1p}, therefore the singularity observed at small values of 
$\tau$ 
has a simple explanation.

 We have computed $<x>_{\tau,T}$ and  used \rf{d3p} to obtain, in this way, 
the solid curve shown in Fig.~\ref{fig1a}.

\section{ Higher order one-dimensional directed abelian algebras.} 
\label{sect4}

 The quadratic DAA discussed in the previous section can be generalized to 
higher orders. 
These algebras define stochastic models with recurrence states (stable 
configurations) containing at most $N-1$ particles per site. 
As we are going to see all the results obtained for the quadratic
case generalize in a trivial way.
 The order $N$ DAA with conservation of the number of particles is defined 
as follows:
\beq\label{3.1}
a_i^N = \sum_{k=0}^{N-1}\alpha_ka_i^ka_{i+1}^{N-k}, \;
(i=1,\ldots,L-1),
\eeq
\beq \label{3.2}
a_L^N = \sum_{k=0}^{N-1} \alpha_ka_L^k,
\eeq
\beq \label{3.3}
\sum_{k=0}^{N-1}\alpha_k = 1, \quad 0 \leq \alpha_k \leq 1.
\eeq
 One recovers the case $N = 2$, taking $\alpha_0 = \mu$. 
The stationary states $\Phi$ of the family of  
 Hamiltonians (see \rf{2}) associated to these algebras, i. e.,  
\beq \label{3.4}
a_i \Phi = 1,
\eeq
have the expression:
\beq \label{3.5}
\Phi = \prod_{i=1}^L \frac{\sum_{m=0}^{N-1}\beta_ma_i^m}{N-\sum_{s=1}^{N-1}
s\alpha_s}
\eeq
where
\beq \label{3.6}
\beta_m = \sum_{k=0}^m \alpha_k.
\eeq

 The average density 
\beq \label{3.7}
\rho = \frac{\sum_{m=1}^{N-1}m\beta_m}{N- \sum_{s=1}^{N-1}s \alpha_s}
\eeq
varies between $(N - 1)/2$ and $N-1$.

 The algebra \rf{3.1} has $N^L$ one-dimensional representations. 
As in  section 2 we restrict ourselves to the simple 
Hamiltonian \rf{16}. Its eigenspectrum 
 can be obtained in the same way as for the $N = 2$ case. 
As compared with the case $N=2$   one has to solve $N$-th order algebraic equations 
instead of quadratic ones. The dynamic critical exponent stays unchanged 
$z = 1$. 

 We present the study of the avalanches  for the case $N=3$ only. 
The generalization to other values of $N$ will become obvious.

 We take $\alpha_0$ and $\alpha_1$ as the parameters of the algebra. 
Using \rf{3.5} 
 we have 
 \beq \label{3.8}
\Phi = \prod_{i=1}^L \frac{a_i^2+(\alpha_0+\alpha_1)a_{i}+\alpha_0}
{1+2\alpha_0+\alpha_1}.
\eeq
We consider only configurations in which on the first site  we have two particles 
and add one more particle on this site. As in the case $N = 2$, 
we have to 
compute the quantity:
\bea \label{3.9}
&&a_i^n\frac{a_i^2+(\alpha_0+\alpha_1)a_{i}+\alpha_0}
{1+2\alpha_0+\alpha_1}= (1+2\alpha_0+\alpha_1)^{-1} \nonumber \\
&&\times [ A_0^{(n)}a_{i+1}^{n+2}   
+(B_1^{(n)}a_i+B_0^{(n)})a_{i+1}^{n+1} \nonumber \\
&&+
(C_2^{(n)}a_i^2+C_1^{(n)}a_i+C_0^{(n)})a_{i+1}^n \nonumber \\ 
&& +(D_2^{(n)}a_{i}^2+D_1^{(n)}a_i)a_{i+1}^{n-1} 
+  
E_2^{(n)}a_i^2a_{i+1}^{n-2} ].
\eea
Multiplying \rf{3.9} by $a_i$, one obtains the following recurrence relations, valid 
for $n\geq 2$,  
\bea \label{3.10}
C_2^{(n+1)} -C_2^{(n)} &=& -\alpha_0(C_2^{(n)} -C_2^{(n-2)}) \nonumber \\
&&-\alpha_1
(C_2^{(n)}-C_2^{(n-1)}) \nonumber \\
D_2^{(n+1)} -D_2^{(n)} &=& -\alpha_0(D_2^{(n)} -D_2^{(n-2)}) \nonumber \\
&&-\alpha_1
(D_2^{(n)}-D_2^{(n-1)}) \nonumber \\
E_2^{(n+1)} -E_2^{(n)} &=& -\alpha_0(E_2^{(n)} -E_2^{(n-2)}) \nonumber \\
&&-\alpha_1
(E_2^{(n)}-E_2^{(n-1)}) 
\eea
\bea \label{3.11}
&& A_0^{(n+1)} = \alpha_0C_2^{(n)},  B_1^{(n+1)}=\alpha_1C_2^{(n)}+\alpha_0C_2^{(n-1)} \nonumber \\
&& B_0^{(n+1)} = \alpha_0D_2^{(n)},  C_1^{(n+1)}=\alpha_1D_2^{(n)}+\alpha_0D_2^{(n-1)} \nonumber \\
&& C_0^{(n+1)} = \alpha_0E_2^{(n)},  D_1^{(n+1)}=\alpha_1E_2^{(n)}+\alpha_0E_2^{(n-1)} 
\eea
with
\bea \label{3.12a}
&& C_2^{(2)} = \frac{1}{1+2\alpha_0+\alpha_1}, D_2^{(2)} = 
\frac{\alpha_0+\alpha_1}{1+2\alpha_0+\alpha_1}, \nonumber \\ 
&&E_2^{(2)} = \frac{\alpha_0}{1+2\alpha_0+\alpha_1}, C_2^{(1)}=D_2^{(1)}=E_2^{(1)}=0.
\eea 
From \rf{3.10}-\rf{3.12a} we obtain
\beq \label{3.12}
D_2^{(n)}=(\alpha_0+\alpha_1)C_2^{(n)}, E_2^{(n)} = \alpha_0C_2^{(n)}.
\eeq

 We notice that \rf{3.12} has $n$-independent solutions valid for large 
$n$:
\beq \label{3.13}
C_2^{(n)} = C.
\eeq
This implies (see \rf{3.11})
\bea \label{3.14}
&& A_0^{(n)} = \alpha_0C, B_0^{(n)}=\alpha_0(\alpha_0+\alpha_1)C, 
C_0^{(n)}=\alpha_0^2C, \nonumber \\
&&B_1^{(n)}=(\alpha_0+\alpha_1)C, C_1^{(n)} = (\alpha_0+\alpha_1)^2C, \nonumber \\
&&D_1^{(n)}=\alpha_0(
\alpha_0+\alpha_1)C.
\eea

 If we are not interested in what happens in the avalanche on the site $i$, 
we take $a_i = 1$ in \rf{3.9}. We substitute the coefficients in \rf{3.9}
 by 
\rf{3.12}-\rf{3.14} and fix the value of $C=(1+2\alpha_0+\alpha_1)^{-1}$ by 
imposing  that the 
sum of the 
coefficients is equal to $1$. We obtain 
\bea \label{3.15}
&&a_i^n\frac{a_i^2+(\alpha_0+\alpha_1)a_i+\alpha_0}{1+2\alpha_0+\alpha_1} 
{\hat{=}} 
(1+2\alpha_0+\alpha_1)^{-2} \nonumber \\
&& \times\left[\alpha_0a_{i+1}^{n+2}+(\alpha_0+\alpha_1)(1+\alpha_0)a_{i+1}^{n+1}\right. 
\nonumber \\ 
&&+(1+\alpha_0^2+(\alpha_0+\alpha_1)^2))a_{i+1}^n \nonumber \\
&&\left. + (\alpha_0+\alpha_1)(1+\alpha_0)a_{i+1}^{n-1}+\alpha_0a_{i+1}^{n-2}\right].
\eea
 This can be interpreted as a random walker which moves in a symmetric way one and two steps 
up and down. We are therefore back 
 to the Maslov Zhang \cite{MZ} universality class. 


\section{ Directed abelian algebras in various geometries.}\label{sect5}

Up to now we have considered DAA on a one-dimensional lattice. DAA can be 
generalized to more dimensions and different lattices. 
For simplicity, in this section, we confine ourselves to quadratic DAA 
which conserve the number of particles. 
\begin{figure}[ht!]
\centering
{\includegraphics[angle=0,scale=0.46]{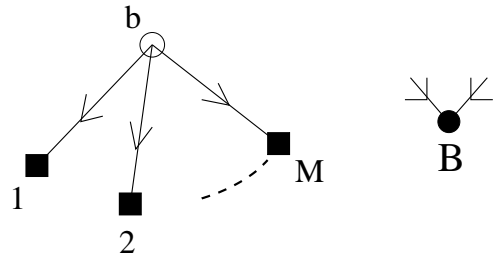}}
\caption{The bulk site $b$ is connected with the sites $1,2,\ldots,M$. 
Bulk sites are denoted by empty circles and boundary sites $B$ by filled 
circles, respectively. Squares denote sites that can be of  bulk or 
boundary type.}
\label{fig2}
\end{figure}

 We divide the sites of a graph  into two sets: bulk sites and 
boundary sites.  Take a bulk site $b$ on the lattice and join it by 
outgoing arrows to the sites $i$ ($i = 1,2,\ldots,M$) (see Fig.~\ref{fig2}). To each of the 
$M + 1$ sites one associate a generator $a_i$. 
\begin{figure}[ht!]
\centering
{\includegraphics[angle=0,scale=0.46]{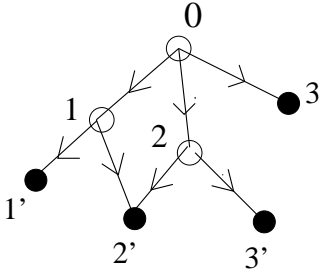}}
\caption{Example of a directed acyclic graph, where the 
empty (filled) circles are bulk (boundary) sites.}
\label{fig3}
\end{figure}
A quadratic directed relation can
be associated to the site $b$:
\beq \label{4.1}
a_b^2 = \sum_{k=1}^M(\alpha_ka_k^2+\beta_ka_ba_k) +
\sum_{k=1}^M\sum_{l=k+1}^M \gamma_{k,l}a_ka_l.
\eeq
 We denote
\beq \label{4.2} \mu = \sum_{k=1}^{N}\alpha_k+ 
\sum_{k=1}^M\sum_{l=k+1}^M\gamma_{k,l}=1-\sum_{k=1}^N\beta_k,
\eeq
where $0<\mu\leq 1$.
 To a boundary (sink) site $B$ (no outgoing arrows) 
one associates a generator 
$a_B$
satisfying the relation:
\beq \label{4.3}
a_B^2 = \mu + (1-\mu)a_B.
\eeq

 The sites $i = 1,2,\ldots,M$ can be bulk or boundary sites to which we associate  
relations like \rf{4.1} or \rf{4.3}, respectively
 (see Fig.~\ref{fig3}). There are as many relations as  sites. 
The graph 
starts with bulk sites and ends with boundary sites. This is a directed 
acyclic graph \cite{DAG}.

\begin{figure}[ht!]
\centering
{\includegraphics[angle=0,scale=0.22]{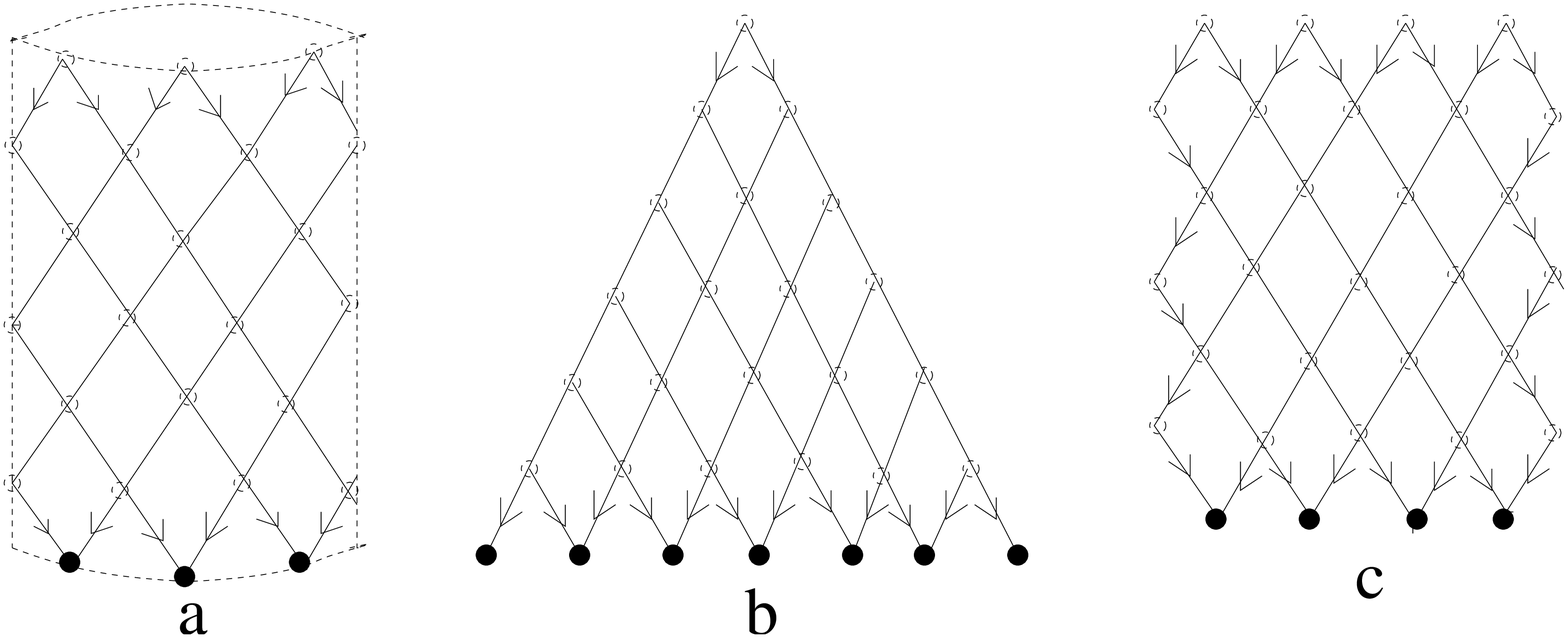}}
\caption{Examples of 2d directed acyclic lattices. In (a) the lattice is periodic 
in the horizontal direction, being defined on a cylinder. In (b) and (c) the lattice is open in the horizontal direction.}
\label{fig4}
\end{figure}

 We call a lattice with $\cal{L}$ sites on which one can define a DAA, a 
directed acyclic lattice. The DAA is defined by the $\cal{L}$ relations \rf{4.1} and 
\rf{4.3}. The algebra has $2^{\cal{L}}$ 
one-dimensional irreducible representations 
and the regular representation is $2^{\cal{L}}$ dimensional. 
Examples of 2-d 
directed acyclic lattices are shown in Fig.~\ref{fig4}. 
If we are interested in critical exponents we have to consider well defined sequences of larger and larger graphs. 

 One can show that for  any directed acyclic lattice one has: 
\beq \label{4.4}
a_I\Phi = \Phi, \quad I\in \cal{L}
\eeq
where
\beq \label{4.5}
\Phi = \prod_{J\in \cal{L}} \frac{\mu + a_J}{1+\mu}.
\eeq
$I$ and $J$ are all the sites of the directed acyclic lattice with $\cal{L}$ sites. 
The 
proof is by finite induction, similar to the one used for the one-dimensional case.
One starts with the boundary generators and finishes  with the bulk generators 
attached to sites with no incoming arrows.

 One can define a simple stochastic process on the directed acyclic lattice  by 
the Hamiltonian
\beq \label{4.6}
H = 1- \frac{1}{\cal{L}} \sum_I a_I,
\eeq
and the $\cal{L}$ quadratic relations \rf{4.1}-\rf{4.3}. 
Obviously the stationary state of 
the system  is given by \rf{4.5}:
\beq \label{4.7}
H\Phi = 0.
\eeq

 In order to find the spectrum of $H$, one uses the diagonal 
representations 
of the generators of $a_I$. The eigenvalues of $a_I$ are obtained using 
the one-dimensional representations of the DAA. One starts with the 
boundary generators (this is the last row in the lattices of Fig.~\ref{fig3}), 
solving the equations \rf{4.3}. The solutions are introduced in the 
quadratic relations \rf{4.1} related to the sites in the next layer (this 
is the row before last  in Fig.~\ref{fig4}). The  layer by layer procedure  is repeated  
(we go against the arrows) up to when we have exhausted the lattice. The 
energy gap corresponding to the first excited state, is obtained taking 
for all the generators the eigenvalue $1$ except for one generator 
belonging to the top of the lattice (there is only one in Fig.~\ref{fig4}b), and many in Figs.~\ref{fig4}a) and \ref{fig4}c)) for 
which we take the eigenvalue $-\mu$. This gives an energy gap:
\beq \label{4.8}
E_1 = (1 + \mu)/{\cal{L}},
\eeq         
and we conclude that the system is always critical and if the 
dimension of the directed acyclic lattice is $D$, the dynamic critical exponent 
is equal to  $z = D$. 

 Before discussing  2-dimensional directed acyclic lattices, we  
consider again a one-dimensional case in which the bulk sites in \rf{4.1} 
are joined to two successive sites. The DAA is:
\bea \label{4.9}
a_i^2 &=&\alpha_1a_{i+1}^2 +\alpha_2a_{i+2}^2+\beta_1a_ia_{i+1} 
+\beta_2a_ia_{i+2} \nonumber \\
&+&\gamma a_{i+1} a_{i+2} \nonumber \\
 a_{L-1}^2 &=&\alpha_2+\beta_2a_{L-1}+\gamma a_L +\alpha_1a_L^2 \nonumber \\
&+&\beta_1a_{L-1}a_L \nonumber \\
a_L^2 &=&\mu +(1-\mu)a_L 
\eea

 We have shown that in general, we spare the reader the effort of having 
to follow a long proof, 
 the avalanches are again given by a random walker. We would like to 
mention only an interesting observation. Let us take all the coefficients 
\rf{4.9} zero except for $\gamma$. We get the deterministic algebra:
\bea \label{4.10}
&& a_i^2 = a_{i+1}a_{i+2} \quad (i=0,1,\ldots,L) \nonumber \\
&& a_{L-1}^2 = a_L,\quad a_L^2 =1 
\eea
and the stationary state is (see \rf{4.5})
\beq \label{4.11}
\Phi = \prod_{i=1}^L\frac{1+a_i}{2}.
\eeq

For simplicity, we take $L + 1$ sites ($i = 0,1,\ldots,L$) with $L$ even and assume that we have two 
particles on the site $i=0$ which trigger the avalanche. 
We also group in the 
ground-state wavefunction \rf{4.11} the sites in pairs. Using \rf{4.10}
 and \rf{4.5} we have 
\bea \label{c1}
&&a_0^2\Phi = a_1a_2\prod_{i=1}^{L/2} (1+a_{2i-1})(1+a_{2i})/4 \nonumber \\
&&=(\frac{a_1a_2}{4}+\frac{a_2a_3a_4}{4}+ \frac{(1+a_2)a_3^2a_4}{4}) 
\nonumber \\
&&\times\prod_{i=2}^{L/2}\frac{(1+a_{2i-1})(1+a_{2i})}{4} 
\hat{=} (\frac{1}{4}+\frac{a_3a_4}{4}+\frac{a_3^2a_4}{4}) \nonumber \\
&&\times \prod_{i=2}^{L/2}\frac{(1+a_{2i-1})(1+a_{2i})}{4}. 
\eea
There are three terms in the last equation. The first one gives 
with a probability $1/4$ an avalanche of size $2$. The second term, with a 
probability $1/4$, has the same structure as the same equation  for a lattice 
with $L - 2$ sites. Using the identity:
\bea \label{60a}
&&a_{2i-1}^2a_{2i}\frac{(1+a_{2i-1})(1+a_{2i})}{4} \nonumber \\  
&&= \frac{(1+a_{2i-1})(1+a_{2i})a_{2i+1}a_{2i+2}}{4}   
\hat{=} a_{2i-1}^2a_{2i+2},
\eea 
we find that the last term in \rf{c1} describes $3$ particles which 
"fly" to 
the boundary and leave the system. This occurs with probability $1/2$. 
Notice (see \rf{4.10}) that $a_{L-1}^2a_{L} = 1$.

 To sum up, if $L$ is large, one gets with a probability $1/4^k$
  an avalanche 
of size $2k$. This implies that with a probability $1/3$ one gets 
finite avalanches. With a probability $2/3$ one gets "avalanches" of the 
size of the system.

 Let us mention that in two dimensions a similar deterministic model  gives avalanches 
with long tails (exponent $\sigma_{\tau} = 3/2$) \cite{DR}. 

We have studied the behavior of avalanches in the two-dimensional 
directed cyclic lattice shown in 
Fig.~\ref{fig4}b using the quadratic DAA 
algebra
\bea \label{dk3}
&&a_{i,j}^2 = \alpha(\mu a_{i+1,j}^2+(1-\mu )a_{i,j}a_{i+1,j}) + 
\nonumber \\
&&(1-\alpha)(\mu a_{i+1,j}^2+(1-\mu )a_{i,j}a_{i+1,j}). 
\eea

 The avalanches observed in the lattice in Fig.~\ref{fig4}b 
coincide with those observed in
the cylinder in Fig.~\ref{fig4}a if the perimeter of the 
cylinder is larger than the size
of the avalanche.

  We did extensive Monte Carlo simulations for several values of the
parameters in the DAA. We took only small values of $\mu$ 
to make sure that
we don't get crossover effects like those described in Fig.~\ref{pp2}. 
We have
considered up to $1.2\times 10^7$ avalanches of sizes up to $T = 30000$. 
It is 
important to note that the estimate is smaller than the expected value.
 
The 
exponent $\sigma_{\tau}$ was obtained in several ways. A possible 
 estimate is the value of $\omega=\sigma_{\tau}$ that makes
\beq \label{est2}
\frac{\partial F(\omega,T)}{\partial T}{\Big |}_{\omega=\sigma_{\tau}} = 0, 
\eeq
for $17,500 <T<30,000$, where 
\beq \label{est3}
F(\omega,T) = T^{\omega-1} \int_{T}^{\infty}p(t)dt.
\eeq
In Fig.~\rf{aa1} we show for $\alpha=1/2$ and $\mu =1/2$ the 
function  
$F(\omega,T)$ for some values of $\omega$. The figure indicate 
$1.77 < \sigma_{\tau} < 1.78$. The same calculation were done for 
$\alpha=1/2, \mu=1/10$ giving us $1.780<\sigma_{\tau}<1.785$, 
and also for the anisotropic lattice where $\alpha=1/4,\mu =1/2$ 
giving us $1.78 <\sigma_{\tau} <1.79$. Our results indicate that the 
model in two 
dimensions belong to a universality class where $\sigma_{\tau} = 1.78 \pm 
0.01$.
We should keep in mind that in one dimension for $\mu = 0.1$, and $T =
30000$, one has obtained an estimate $\sigma_{\tau} (T) = 1.49971$ , quite close to and
smaller than the correct value 3/2.
 Previous Monte-Carlo simulations \cite{RV} had larger errors. If one takes our
results at face value 
 they contradict the theoretical predictions \cite{PB,KMY}. We cannot exclude however that larger values of $T$ are necessary to get the ultimate 
value of $\sigma_{\tau}$.
%
\begin{figure}[ht!]
\centering
{\includegraphics[angle=0,scale=0.46]{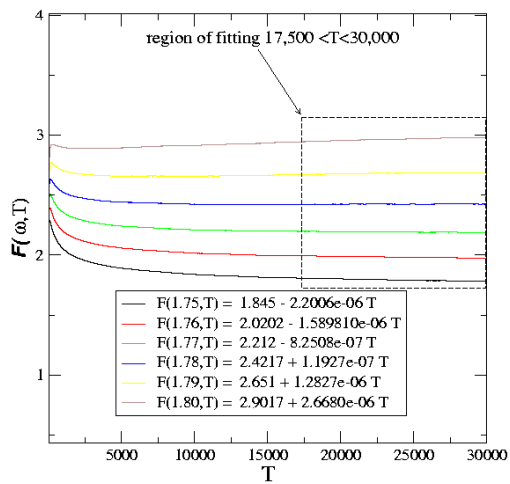}}
\caption{(Color online) 
The function $F(\omega,T)$ as defined in \rf{est3} for 
some values of $\omega$. This figure is for the stochastic model associated  with \rf{dk3} with $\alpha=1/2$ and $\mu =1/10$. In the inset we show the linear fit for the values of $\omega$ obtained from the 
datas $17,500<T<30,000$.} 
\label{aa1}
\end{figure}

\section{Conclusions}

Using directed acyclic lattices \cite{DAG} we have defined directed abelian 
algebras (DAA). The lattices can be in any dimensions. We attach a 
generator of the DAA to each site of the lattice. There are polynomial  
relations among the generators. A given DAA can depend on several 
parameters. The remarkable property of these algebras is that they are
in general semisimple.

 For each DAA one can define a family of Hamiltonians acting onto the 
regular representation of the DAA. These Hamiltonians give the time 
evolution of various stochastic processes. The vector space of the regular 
representation can be mapped in the vector space space obtained taking $N$ 
states on each site of the directed acyclic lattice ($N$ is the degree of the polynomial
relations). In this way one obtains $N$-state Hamiltonians in an arbitrary
number of dimensions.

 It is easy to compute the whole spectrum for the family of Hamiltonians 
related to a given DAA. All one has to do is to find all the 
one-dimensional irreducible representations of the algebra. This is a 
purely algebraic exercise. Once the spectrum of the generators in the 
diagonal representation is known one can compute the spectrum of the 
Hamiltonians.

 To find the eigenfunctions of the Hamiltonians, one has to solve 
a purely algebraic problem: one writes  the one-dimensional 
representations in the basis of the regular representation. To find the 
ground state wavefunctions corresponding to the stationary states one has 
to consider the irrep  in which all the generators are equal to one. The 
stationary states can easily be obtained for any Hamiltonian obtained from 
a DAA. The stationary states always have  a product measure (there are no 
correlations) and depend on fewer parameters than the DAA. All the
wavefunctions can probably be obtained. In a simple example (see Appendix 
A) we show not only how to obtain the whole spectrum but give also all the 
wavefunctions. They have a simple factorized form.  

 We have studied some simple Hamiltonians in which only the generators 
appear. For a simple mathematical reason, in the finite-size scaling 
limit, they are all gapless with a dynamic critical exponent $z = D$ ($D$
 is the
number of dimensions of the lattice). There are many interesting 
time dependent physical properties which are worth studying in these 
models, the simplest being the time evolution of local densities. We plan 
to look at these aspects of the models in the future. Instead we have 
studied the properties of avalanches.
 Each DAA together with the ground state wavefunction of the Hamiltonian 
can define avalanches. 

In  one dimension  and in a DAA 
algebra which conserves the number of particles, we have checked for two 
and three-state models that one gets avalanches in the "random walk" 
universality class \cite{MZ}. We have also studied the "composition" of 
avalanches. One has a depletion of particles at the position where the 
avalanche starts and an enrichment at the end of the avalanche.
 The Hamiltonian is gapless even if the DAA does not conserve the number 
of particles. This implies that one doesn't have SOC (the probability to 
have large avalanches decreases exponentially) in spite of the fact that 
the Hamiltonian is critical. One can complement this observation with 
another similar observation. Since we are in one dimension the stationary
 state of a $N$-state model can 
be seen as a spin model defined on the dual lattice. The spin model 
depends on many parameters (there is only one for $N = 2$ corresponding to 
the temperature) and is not critical in spite of having a stochastic model 
which is critical. The opposite situation is already known \cite{ADMR}: a 
Hamiltonian having a gap drives the system to the critical (zero 
temperature) spin system.

 In two dimensions all we did is to check that for various 
choices of parameter one stays in the universality class described in 
\cite{RV,PB,KMY}.

\section*{Acknowledgments}
G. M. Nakamura has participated in the initial stage of this work, and we 
thank him for further discussions.
We thank V.B. Priezzhev and D. Dhar for fruitful discussions, and to A. Owczarek for providing us with the relation \rf{d4p}. 
This work has been partially supported by the Brazilian agencies FAPESP and 
CNPq.

\appendix

\section{
Spectrum and eigenfunctions of a Hamiltonian derived from a quadratic 
directed abelian algebra}

In section 2 and 3 we have discussed the quadratic abelian algebra
\bea \label{a.1}
a_i^2 =&& \mu a_{i+1}^2+(1-\mu)a_ia_{i+1},\quad i=1,2,\ldots,L, \nonumber \\
&&a_{L+1}=1, \quad 
[a_i,a_j]=0,
\eea
and the Hamiltonian:
\beq \label{a.2}
H = \frac{1}{L}\sum_{i=1}^L (1-a_i).
\eeq

 In order to obtain the spectrum of $H$ we have to find 
the representations of $a_i$ in the diagonal representation. In order to 
find the wavefunctions one has to express the $a_i$ in the regular 
representation. In the latter representation, the vector space is given by
monomials of the $a_j$'s (see section 2).  

 In the representation in which $a_i$ are diagonal, 
the distinct eigenvalues 
$x_i$ of $a_i$ can be obtained recursively from the eigenvalues $x_{i+1}$
 of $a_{i+1}$:
\beq \label{a.3}
(x_i-x_{i+1})(x_i+\mu x_{i+1}) = 0.
\eeq
 Equation \rf{a.3} can be written as:
\beq \label{a.4}
x_i = (-\mu)^{\epsilon_i}x_{i+1} \quad (\epsilon_i=0,1).
\eeq
 This implies that the  eigenvalues of $a_{L-l}$ are:
\beq \label{a.5}
x_{L-l}(\epsilon_{L-l},\epsilon_{L-(l-1)},\ldots,\epsilon_L) = 
\prod_{j=0}^l (-\mu)^{\epsilon_{L-j}},
\eeq
and are labelled by the $\epsilon_i = 0, 1$ ($i = L, L-1,\ldots,L-l$).
 In the $2^L$ representation of the algebra \rf{a.1}, the eigenvalues of 
$a_{L-l}$ are:
\bea \label{a.6}
&&x_{L-l}(\epsilon_1,\epsilon_2,\ldots,\epsilon_{L-l-1},\epsilon_{L-l},
\epsilon_L)  \nonumber \\
&&=\prod_{k=1}^{L-l-1}(1)^{\epsilon_k} \prod_{j=0}^l 
(-\mu)^{\epsilon_{L-j}},
\eea
except for $a_1$, they are degenerate.

 We are going to show that in the regular representation, the factorized 
expression
\bea \label{a.7}
&&\Psi_(\epsilon_{L-l},\epsilon_{L-l+1},\ldots,\epsilon_L) 
\nonumber \\
=\prod_{m=0}^l&&\left[\mu + (-\mu)^{\epsilon_{L-m}-\sum_{j=0}^{m-1}\epsilon_{L-j}} 
a_{L-m}\right]
\eea
is an eigenfunction of $a_{L-l}$ corresponding to the eigenvalue \rf{a.5}:
\bea \label{a.8}
&&a_{L-l}
\Psi(\epsilon_{L-l},\ldots,\epsilon_{L-1},\epsilon_L)
 = x_{L-l}(\epsilon_{L-l},\ldots,\epsilon_{L-1},\epsilon_L)\nonumber \\ 
&& \times \Psi(\epsilon_{L-l},\ldots,\epsilon_{L-1},\epsilon_L).
\eea

 This implies that $\Psi_{\epsilon_1,\ldots,\epsilon_L}$ is an 
eigenfunction of 
all $a_i$ ($i =1,\ldots,L$) corresponding to the eigenvalues \rf{a.6}. In 
particular if we take all the $\epsilon_i = 0$, one obtains the eigenvalue 
one for all the $a_i$ which gives an eigenvalue zero for $H$  (see 
\rf{a.2}) and the corresponding eigenfunction is, up to a normalization factor,
equal to \rf{12}.
 
The spectrum of $H$ is obtained introducing the eigenvalues \rf{a.6} in 
\rf{a.2}.

Obviously \rf{a.7} is also an eigenfunction for all $a_{L-j}$ ($0<j<l$).
 
 Using \rf{a.6} one can compute all the eigenvalues of $H$ 
(they are fixed by 
the values of the $\epsilon_i, i = L, L-1,\ldots,1$). The corresponding 
eigenfunctions being given by \rf{a.7} with $l = L - 1$.
 The proof of \rf{a.8} is by finite induction.

a) \rf{a.8} is valid for $l = 0$.
 Using \rf{a.1} one has:
\beq \label{a.9}
a_L(\mu +(-\mu)^{\epsilon_L}a_L) = 
(-\mu)^{\epsilon_L}(\mu + (-\mu)^{\epsilon_L}a_L),
\eeq
in agreement with \rf{a.8}.

b) If \rf{a.8} is valid for $l$, it is valid for $l+1$
 We have to check that $a_{L-l-1}$ applied to 
$\Psi(\epsilon_{L-l-1},\ldots)$  
gives the eigenvalues \rf{a.6}. One has:
\bea \label{a.10}
&&a_{L-l-1}  
\Psi(\epsilon_{L-l-1},\epsilon_{L-l},\ldots,\epsilon_L)\nonumber \\
&&= a_{L-l-1}
(\mu +(-\mu)^{\epsilon_{L-l-1}-\sum_{j=0}^l\epsilon_{L-j}}
a_{L-l-1}) \nonumber \\ 
&&\times \Psi(\epsilon_{L-l},\epsilon_{L-l+1},\ldots,\epsilon_L)=
x_{L-l-1}(\epsilon_1,\epsilon_2,\ldots,\epsilon_L)\nonumber \\ 
&&\times \Psi(\epsilon_{L-l-1},\epsilon_{L-l},\ldots,\epsilon_L),
\eea
where we have used \rf{a.1} and \rf{a.8}.

 We think that the factorized form of the eigenfunctions \rf{a.7} is a 
general property  of stochastic models obtained from DAA. It is 
a consequence of the representation theory of the DAA and not of a 
specific form of the Hamiltonians.

\section{ Several representations of the quadratic DAA algebra (5)}

 We have to stress that unlike the spectra of evolution operators 
which are insensitive to the choice of the representations of the 
DAA algebras, the physical properties  of the systems depend 
on the choice of the representation. For example, the existence or not of 
avalanches or even the universality classes of avalanches are not 
determined  by the 
algebras but by their representations. Similarly a redefinition of the 
linearly independent words $W(r)$ (see Eq.~\rf{1}) in the algebra 
changes the 
physics. For example, the shift $a_i \in  a_i' + \gamma$ in the algebra \rf{5} can make
one of the coeficients $p_2, p_3$ or $q_2$ vanish, such a shift changes the 
physical process.   

 In order to clarify the importance of choosing different representations 
for the same algebra, we consider the simpler case of the quadratic 
algebra \rf{22}:
\bea \label{b1}
&&a_i^2 = \mu a_{i+1}^2 + (1-\mu)a_i a_{i+1}  (i = 1,2,\ldots L), 
\nonumber \\
&& a_i a_j =a_j a_i, \quad a_{L+1} = 1.     
\eea

 The expression of the generators in the $2^L$ regular representation 
used in 
Sections 2 and 3 can be written in the following way: 
\beq \label{b2}
a_i = I_{1,i-1} \otimes A_i,                                       
\eeq
where $I_{i,j}$  is the $2^{j-i}$ unit matrix acting onto the tensor 
product 
$V_i\otimes V_(i+1)\otimes \cdots \otimes V_(j)$, $V$ is a 2-dimensional 
vector space. $A_i$ is a $2\times 2$ 
matrix, with elements acting on a $2^{L-i}$ vector space (order $1$ and $a_i$)
\bea \label{b3}
&&A_i =  
\left(\begin{array}{cc} 
  0 &  \mu A_{i+1}^2 \\
  I_{i+1,L} & (1-\mu)A_{i+1}  \end{array}\right) 
\eea

 A second representation of the algebra is:
\beq \label{b4}
a_i = I_{1,i-1}\otimes B_i                                        
\eeq
where 
\bea \label{b5}
&& B_i = \left(\begin{array}{cc} 
1-\alpha & \beta \\
\alpha & 1-\beta \end{array} \right)\otimes B_{i+1},
\eea
with 
\beq \label{b6}
 \mu = \alpha +  \beta - 1.
\eeq                                   
There are no avalanches in this representation. A particle hitting the 
site $i$ doesn't influence the site $i + 1$.

 A third representation is used in the totally asymmetric Oslo model 
\cite{SC}:
\bea \label{b7}
&&a_i = I_{1,i-1}\otimes C_i
\eea                                        
where
\bea \label{b8}
C_i =  \left( \begin{array}{cc} (1- \alpha)C_{i+1} & \beta C_{i+1}^2 \\
          \alpha I_{i+1,L} &           (1-\beta)C_{i+1}\end{array} 
\right) 
\eea
with
$\mu = \alpha +  \beta - 1$.

 The eigenvalues of $a_i$ are the same for all the three representations. 
The eigenfunction $\Phi$
\beq \label{b9}
a_i \Phi = \Phi \quad (i = 1,2,\ldots ,L)                            
\eeq
is also the same:
\beq 
\Phi = \bigotimes_{i=1}^L  \frac{1}{\alpha +\beta} \left(\begin{array}{c} 
\beta \\ \alpha\end{array}\right)_i.
\eeq

For the representation \rf{b3} one has to take $\alpha = 1, \beta = \mu$ in \rf{b9} 
and obtain \rf{12} which was used in Sections 2 and 3.

 One can repeat the calculations done in Section 3 and derive the
properties of avalanches. The results are a simple generalization of the
previous results. Equation 
\rf{e1} stays unchanged with $\mu$ given by \rf{b6}
 and
the diffusion constant is $D = \alpha \beta/(\alpha + \beta)^2$. For $\alpha = 1$,
one recovers the results of Section 3. The parameter $\mu$ in the algebra 
\rf{b1}
 can take negative values. This is not the case if $\alpha = 1$. The
average density in the stationary state
\beq \label{b11}
\rho = \alpha/(\alpha + \beta)   
\eeq
spans the interval $0 \leq \rho \leq 1$. The interval is smaller if $\alpha = 1$.


\begin{thebibliography}{99}
\bibitem{footnote}A left $I$ ideal of an algebra $A$ is a subspace of $A
$ such that $aX \in I$ for all
$a \in A$ and $X \in I$.
\end{thebibliography}
\end{document}